\documentclass{JHEP3}

\newcommand {\nn}    {\nonumber}

\title{Localization of Fermions on a String-like Defect}

\author{Yu-Xiao Liu, Li Zhao and Yi-Shi Duan \\
 Institute of Theoretical Physics, Lanzhou University\\
 Lanzhou, 730000, P. R. China\\
 E-mail: \email{liuyx@lzu.edu.cn}, \email{zhl03@lzu.cn}, \email{ysduan@lzu.edu.cn} }

\abstract{We study localization of bulk fermions on a string-like
defect with the exponentially decreasing warp factor in six
dimensions with inclusion of U(1) gauge background from the
viewpoint of field theory, and give the conditions under which
localized spin 1/2 and 3/2 fermions can be obtained.}


\begin{document}

\section{Introduction}

Recently, there has been considerable activity in the study of
models that involve new extra dimensions. The possible existence
of such dimensions got strong motivation from theories that try to
incorporate gravity and  gauge interactions in a unique scheme, in
a reliable manner. The idea dates back to the 1920's,  to the
works of Kaluza and Klein \cite{KaluzaKlein} who tried to unify
electromagnetism with Einstein gravity by assuming that the photon
originates from the fifth component of the metric.

Suggestions that extra dimensions may not be compact
\cite{RubakovPLB1983136}-\cite{RS} or large
\cite{AntoniadisPLB1990,ADD} can provide new insights for a
solution of gauge hierarchy problem \cite{ADD}, of cosmological
constant problem
\cite{RubakovPLB1983139,Randjbar-DaemiPLB1986,KehagiasPLB2004},
and give new possibilities  for model building. In Ref. \cite{RS},
an alternative scenario of the compactification has been put
forward. This new idea is based on the possibility that our world
is a three brane embedded in a higher dimensional space-time with
non-factorizable warped geometry. In this scenario, we are free
from the moduli stabilization problem in the sense that the
internal manifold is noncompact and does not need to be
compactified to the Planck scale any more, which is one of reasons
why this new compactification scenario has attracted so much
attention. An important ingredient of this scenario is that all
the matter fields are thought of as confined to a 3-brane, whereas
gravity is free to propagate in the extra dimensions.


Following the brane world models proposed by Randall and Sundrum
\cite{RS}, a fair amount of activity has been generated involving
possible extensions and generalizations, among which,
co--dimension two models in six dimensions have been a topic of
increasing interest
\cite{6Dmodels,GherghettaPRL2000,OdaPLB2000113,6dmodel}. A useful
review on topological defects in higher dimensional models and its
relation to braneworlds is available in {\cite{roessl}}. Apart
from model construction, the question of solving the cosmological
constant problem has been the primary issue addressed in several
articles \cite{cosmoconstant6d}. Other aspects such as cosmology,
brane gravity {\cite{codim2gen}, fermion families and chirality
{\cite{ErdemEPJC2002} etc. have been discussed by numerous
authors. A list of some recent articles on codimension two models
is provided in \cite{codim2recent}.

It is well--known by now that in the braneworld scenario it is
necessary to introduce dynamics which can determine the location
of the branes in the bulk.
Ever since Goldberger and Wise \cite{goldwise} added a bulk scalar
field to fix the location of the branes in five dimensions,
investigations with bulk fields became an active area of research.
It has been shown that the graviton \cite{RS} and the massless
scalar field \cite{BajcPLB2000} have normalizable zero modes on
branes of different types, that the Abelian vector fields are not
localized in the Randall-Sundrum (RS) model in five dimensions but
can be localized in some higher-dimensional generalizations of it
\cite{OdaPLB2000113}. In contrast, in
\cite{BajcPLB2000,NonLocalizedFermion} it was shown that fermions
do not have normalizable zero modes in five dimensions, while in
\cite{OdaPLB2000113} the same result was derived for a
compactification on a string \cite{GherghettaPRL2000} in six
dimensions. Subsequently, Randjbar-Daemi {\em et al} studied
localization of bulk fermions on a brane with inclusion of scalar
backgrounds \cite{RandjbarPLB2000} and minimal gauged supergravity
\cite{Parameswaran0608074} in higher dimensions and gave the
conditions under which localized chiral fermions can be obtained.

Since spin half fields can not be localized on the brane
\cite{RS,OdaPLB2000113} in five or six dimensions by gravitational
interaction only, it becomes necessary to introduce additional
non-gravitational interactions to get spinor fields confined to
the brane or string-like defect. The aim of the present article is
to study localization of bulk fermions on a string-like defect
with codimension 2 in U(1) gauge background. The solutions to
Einstein's equations in two extra dimensions have been studied by
many groups
\cite{KehagiasPLB2004,6Dmodels,GherghettaPRL2000,Vilenkin,EinsteinEq}.
In this article, we first review the solutions with a warp factor
in a general space-time dimension. Then, we shall prove that spin
1/2 and 3/2 fields can be localized on a defect with the
exponentially decreasing warp factor if gauge and gravitational
backgrounds are considered.

\section{A string-like defect}
Let us start with a brief review of a string-like defect solution
to Einstein's equations with sources. We consider Einstein's
equations with a bulk cosmological constant $\Lambda$ and an
energy-momentum tensor $T_{MN}$ in general $D$ dimensions:
\begin{eqnarray}
R_{MN} - \frac{1}{2} g_{MN} R = - \Lambda g_{MN}  + \kappa_D^2
T_{MN}, \label{EinsteinEq}
\end{eqnarray}
where $\kappa_D$ denotes the $D$-dimensional gravitational
constant  with a relation $\kappa_D^2 = 8 \pi G_N = {8 \pi}/
{M_*^{D-2}}$, $G_N$ and $M_*$ being the $D$-dimensional Newton
constant and the $D$-dimensional Planck mass scale, respectively,
the energy-momentum tensor is defined as
\begin{eqnarray}
T_{MN} = - \frac{2}{\sqrt{-g}} \frac{\delta}{\delta g^{MN}} \int
d^D x \sqrt{-g} L_m.
\end{eqnarray}
Throughout this article we follow the standard conventions and
notations of the textbook of Misner, Thorne and Wheeler
\cite{Misner}.

We shall consider $D = (D_1+D_2+1)$-dimensional manifolds with the
geometry
\begin{eqnarray}
ds^2 &=& g_{MN} dx^M dx^N  \nn\\
&=& \mathrm{e}^{-A(r)} \hat{g}_{\mu\nu}(x) dx^\mu dx^\nu +
\mathrm{e}^{-B(r)} \tilde{g}_{ab}(y) dy^{a} dy^{b} + dr^2 ,
\label{ds2}
\end{eqnarray}
where $M, N$ denote $D$-dimensional space-time indices,
$\mu,\nu=0,1,\dots,D_1-1, \quad a,b=1,\dots,D_2$, and the
coordinates $y^a$ cover an internal manifold $K$ with the metric
$\tilde{g}_{ab}(y)$. Moreover, we shall adopt the ansatz for the
energy-momentum tensor respecting the spherical symmetry:
\begin{eqnarray}
  T^\mu_\nu = \delta^\mu_\nu t_1(r),  ~~
  T^a_b = \delta^a_b t_2(r), ~~
  T^r_r = t_3(r),  \label{TMN}
\end{eqnarray}
where $t_i(i=1, 2, 3)$ are functions of only the radial coordinate
$r$.

Under these ansatzs, Einstein's equations (\ref{EinsteinEq}) and
the conservation law for energy-momentum tensor $\nabla^M T_{MN} =
0$ reduce to
\begin{eqnarray}
 \mathrm{e}^A \hat{R} + \mathrm{e}^B \tilde{R}
      - \frac{1}{4} D_1(D_1-1) (A')^2
      - \frac{1}{4} D_2(D_2-1)(B')^2 ~~~~~~~~~~~~~~~~~~\nn\\
      -\frac{1}{2}D_1 D_2 A' B'
   - 2\Lambda + 2 \kappa_D^2 t_3 = 0,\\
 \mathrm{e}^A \hat{R} + \frac{D_2 -2}{D_2}~ \mathrm{e}^B \tilde{R} + D_1 A''
             + (D_2-1) B''- \frac{1}{2}D_1(D_2-1) A' B'~~~~~~~~~~~~~~~~~~ \nn\\
     - \frac{1}{4}D_1(D_1 +1) (A')^2
     - \frac{1}{4}D_2(D_2 +1) (B')^2
     - 2\Lambda + 2 \kappa_D^2 t_2 = 0,\\
 \mathrm{e}^B \tilde{R} + \frac{D_1 -2}{D_1}~ \mathrm{e}^A \hat{R} + D_2 B''
             + (D_1-1) A''- \frac{1}{2}D_2(D_1-1) A' B' ~~~~~~~~~~~~~~~~~~\nn\\
     - \frac{1}{4}D_1(D_1 +1) (A')^2
     - \frac{1}{4}D_2(D_2 +1) (B')^2
     - 2\Lambda + 2 \kappa_D^2 t_1 = 0,\\
 t'_3 = \frac{1}{2}D_1 A' (t_3 - t_1)
  + \frac{1}{2}D_2 B' (t_3 - t_2),~~~~~~~~~~~~~~~~~~~~~~~
\end{eqnarray}
where $\hat{R}$ and $\tilde{R}$ are the scalar curvatures
associated with the metric $\hat{g}_{\mu\nu}$ and
$\tilde{g}_{ab}$, respectively, and the prime denotes the
derivative with respect to $r$. Here we define the cosmological
constant $\hat\Lambda$ on the $(D_1-1)$-brane by the equation
\begin{eqnarray}
 \hat{R}_{\mu\nu} - \frac{1}{2} \hat{g}_{\mu\nu} \hat{R}
   = - \hat\Lambda \hat{g}_{\mu\nu}.
\end{eqnarray}

It is now known that there are many interesting solutions to these
equations (see, for instance, \cite{Vilenkin}). Here, we shall
confine ourselves to the brane solutions with a warp factor
\begin{eqnarray}
 A(r) = c r, \label{12}
\end{eqnarray}
where $c$ is a constant.

If $K$ is taken as a $D_2$-torus, then we have $\tilde{R}=0$, and
the general solutions with the warp factor (\ref{12}) can be found
as follows:
\begin{eqnarray}
 ds^2 = \mathrm{e}^{-cr} \hat{g}_{\mu\nu} dx^\mu dx^\nu
   + dr^2 + R_{0}^2 ~\mathrm{e}^{-B(r)} \delta_{ij} d\theta^i d\theta^j, \label{SolutionOfTorus}
\end{eqnarray}
where
\begin{eqnarray}
 B(r) &=& cr + \frac{4}{D_1 c} \kappa_D^2 \int^r dr (t_3 - t_2), \\
 c^2 &=& \frac{1}{D_1(D_1+1)}(-8 \Lambda + 8 \kappa_D^2 \alpha), \\
 \hat{R} &=& \frac{2D_1}{D_1-2} \hat\Lambda = -2 \kappa_D^2 \beta.
\end{eqnarray}
Here $t_2$ takes the following form
\begin{eqnarray}
t_2 = \alpha + \beta \mathrm{e}^{cr}, \label{27}
\end{eqnarray}
with $\alpha$ and $\beta$ being some constants. Moreover, in order
to guarantee the positivity of $c^2$, $\alpha$ should satisfy an
inequality $-8 \Lambda + 8 \kappa_D^2 \alpha > 0$.

If $K$ is taken as a unit $D_2$-sphere, then we have
\begin{eqnarray}
 d \Omega_{D_2}^2
 &=& \tilde{g}_{ab}(y) dy^{a} dy^{b} \nn\\
 &=& d\theta_1^2 + \sin^2 \theta_1 d\theta_2^2
   + \sin^2 \theta_1 \sin^2 \theta_2 d\theta_3^2 + \cdots + \prod_{i=1}^{D_2 -1} \sin^2 \theta_i d\theta_{D_2}^2. \label{5}
\end{eqnarray}
In the case of $D_2=1$, we have $\tilde{R}=0$ and the solutions
are the same as those of (\ref{SolutionOfTorus})
\cite{OdaPLB2000113}. For $D_2 \ge 2$ the solution with the warp
factor (\ref{12}) is of the form \cite{OdaPLB2000113}
\begin{eqnarray}
ds^2 = \mathrm{e}^{-cr} \hat{g}_{\mu\nu} dx^\mu dx^\nu + dr^2 +
R_0^2 d \Omega_{D_2}^2, \label{32}
\end{eqnarray}
where
\begin{eqnarray}
c^2 &=& \frac{-8 \Lambda}{D_1(D_1+D_2-1)}, \\
\hat{R} &=& \frac{2D_1}{D_1-2} \tilde\Lambda = 0, \label{33}
\end{eqnarray}
here the sources satisfy the relations, $t_3 + D_2 t_2 -(D_2-1)t_1
= 0$ and $t_3 = t_1 = constant$, which are nothing but the
relations satisfied in the spontaneous symmetry breakdown
\cite{Vilenkin}.

It is useful to consider a special case of the above general
solutions (\ref{SolutionOfTorus}) with $D_2 =1$. A specific
solution occurs when we have the spontaneous symmetry breakdown
$t_3 = -t_2$ \cite{Vilenkin}:
\begin{eqnarray}
ds^2 = \mathrm{e}^{-cr} \hat{g}_{\mu\nu} dx^\mu dx^\nu + dr^2 +
R_0^2 ~ \mathrm{e}^{-c_1 r} d\theta^2, \label{StringLikeSolution}
\end{eqnarray}
where
\begin{eqnarray}
 c^2 &=& \frac{1}{D_1(D_1 + 1)}(-8 \Lambda + 8 \kappa_D^2 t_2) > 0, \\
 c_1 &=& c - \frac{8}{D_1 c} \kappa_D^2 t_2, \\
 \hat{R} &=& \frac{2D_1}{D_1-2} \hat{\Lambda} = 0. \label{31}
\end{eqnarray}
This special solution would be utilized to analyse localization of
fermionic fields on a string-like defect in the next section.

\section{Localization of fermions}

In this section, for clarity we shall limit our attention to a
specific string-like solution (\ref{StringLikeSolution}) as well
as $D=6$ since the generalization to the general solutions
(\ref{SolutionOfTorus}) is straightforward. In this paper, we have
the physical setup in mind such that `local cosmic string' sits at
the origin $r=0$ and then ask the question of whether various bulk
fermions with spin 1/2 and 3/2 can be localized on the brane with
the exponentially decreasing warp factor by means of the
gravitational interaction and gauge background. Of course, we have
implicitly assumed that various bulk fields considered below make
little contribution to the bulk energy so that the solution
(\ref{StringLikeSolution}) remains valid even in the presence of
bulk fields.

\subsection{Spin 1/2 fermionic field}\label{A}

In this subsection we study localization of a spin 1/2 fermionic
field in gravity (\ref{StringLikeSolution}) and gauge backgrounds.
It will be shown that provided that the gauge field $A_r$
satisfies certain condition, there is a localized zero mode on the
string-like defect.

Let us consider the Dirac action of a massless spin 1/2 fermion
coupled to gravity and gauge field:
\begin{eqnarray}
S_m = \int d^D x \sqrt{-g} \bar{\Psi} i \Gamma^M D_M \Psi,
\label{DiracAction}
\end{eqnarray}
from which the equation of motion is given by
\begin{eqnarray}
\Gamma^M  ( \partial_M + \omega_M  -i e A_M) \Psi=0,
\label{DiracEq1}
\end{eqnarray}
where $\omega_M= \frac{1}{4} \omega_M^{\bar{M} \bar{N}}
\Gamma_{\bar{M}} \Gamma_{\bar{N}}$ is the spin connection with
$\bar{M}, \bar{N}, \cdots$ denoting the local Lorentz indices,
$\Gamma^M$ and $\Gamma^{\bar{M}}$ are the curved gamma matrices
and the flat gamma ones, respectively, and $A_M$ is a U(1) gauge
field. The RS model is the special case with $D_2=0$ and $A_M=0$.
From the formula $\Gamma^M = e^M _{\bar{M}} \Gamma^{\bar{M}}$ with
$e_M ^{\bar{M}}$ being the vielbein, we have the relations:
\begin{eqnarray}
 \Gamma^\mu = \mathrm{e}^{\frac{1}{2}cr}
              \hat{e}^{\mu}_{\bar{\mu}} \Gamma^{\bar{\mu}},~~
 \Gamma^r   = \delta^r_{\bar{r}} \Gamma^{\bar{r}}, ~~
 \Gamma^\theta =R_0^{-1} \mathrm{e}^{\frac{1}{2}c_{1}r} \delta^\theta_{\bar{\theta}}
             \Gamma^{\bar{\theta}}. \label{54}
\end{eqnarray}
The spin connection $\omega_M^{\bar{M} \bar{N}}$ in the covariant
derivative $D_M \Psi = (\partial_M + \frac{1}{4} \omega_M^{\bar{M}
\bar{N}} \Gamma_{\bar{M}} \Gamma_{\bar{N}} -ie A_M ) \Psi$ is
defined as
\begin{equation}
 \omega_M ^{\bar{M} \bar{N}}
 = \frac{1}{2} {e}^{N \bar{M}}(\partial_M e_N ^{\bar{N}}
                      - \partial_N e_M ^{\bar{N}})
     - \frac{1}{2} {e}^{N \bar{N}}(\partial_M e_N ^{\bar{M}}
                      - \partial_N e_M ^{\bar{M}})
     - \frac{1}{2} {e}^{P \bar{M}} {e}^{Q \bar{N}} (\partial_P e_{Q
{\bar{R}}} - \partial_Q e_{P {\bar{R}}}) {e}^{\bar{R}} _M.
\end{equation}
So the  non-vanishing components of $\omega_M$ are
\begin{eqnarray}
  \omega_\mu &=& \frac{1}{4}c \Gamma_r\Gamma_\mu
                + \hat{\omega}_\mu~, \label{eq4} \\
  \omega_\theta &=& \frac{1}{4} c_1 \Gamma_r
            \Gamma_\theta ~, \label{eq5}
\end{eqnarray}
where $\hat{\omega}_\mu=\frac{1}{4} \bar\omega_\mu^{\bar{\mu}
\bar{\nu}} \Gamma_{\bar{\mu}} \Gamma_{\bar{\nu}}$ is the spin
connection derived from the metric
$\hat{g}_{\mu\nu}(x)=\hat{e}_{\mu}^{\bar{\mu}}
\hat{e}_{\nu}^{\bar{\nu}}\eta_{\bar{\mu}\bar{\nu}}$.

Assume $A_{\mu}=A_{\mu}(x)$ and $A_{r,\theta}=A_{r,\theta}(r)$.
The Dirac equation (\ref{DiracEq1}) then becomes
\begin{equation}
 \left\{ \mathrm{e}^{\frac{1}{2}cr} \hat{e}^{\mu}_{\bar{\mu}}
         \Gamma^{\bar{\mu}} \hat{D}_{\mu}
        +\Gamma^r \left( \partial_r - c
                         - \frac{1}{4}c_{1} -ie A_r(r) \right)
        +\Gamma^{\theta}(\partial_\theta-ie A_{\theta}(r)) \right \} \Psi=0, \label{DiracEq2}
\end{equation}
where $\hat{e}^{\mu}_{\bar{\mu}} \Gamma^{\bar{\mu}}
\hat{D}_{\mu}=\hat{e}^{\mu}_{\bar{\mu}} \Gamma^{\bar{\mu}}
({\partial_\mu+\hat\omega_{\mu}-ieA_{\mu}})$ is the Dirac operator
on the brane in the background of the gauge field $A_\mu$. We are
now ready to study the above Dirac equation for 6-dimensional
fluctuations, and write it in terms of 4-dimensional effective
fields. Since $\Psi$ is a 6-dimensional Weyl spinor we can
represent it by \cite{Parameswaran0608074}
\begin{equation}
 \Psi=\left(%
\begin{array}{c}
  \Psi^{(4)} \\
  0 \\
\end{array}%
\right),
\end{equation}
where $\Psi^{(4)}$ is a 4-dimensional Dirac spinor. Our choice for
the 6-dimensional constant gamma matrices $\Gamma^{\bar{M}}$,
$M=0,1,2,3,\bar{r},\bar{\theta}$ are
\begin{equation}
\Gamma^{\bar{\mu}}=
\left(%
\begin{array}{cc}
  0 & \gamma^{\bar{\mu}} \\
  \gamma^{\bar{\mu}} & 0 \\
\end{array}%
\right),~~
\Gamma^{\bar{r}}=
\left(%
\begin{array}{cc}
  0 & \gamma^{5} \\
  \gamma^{5} & 0 \\
\end{array}%
\right),~~
\Gamma^{\bar{\theta}}=
\left(%
\begin{array}{cc}
  0 & -i \\
  i & 0 \\
\end{array}%
\right), \label{Gamma}
\end{equation}
where the $\gamma^{\bar{\mu}}$ are the 4-dimensional constant
gamma matrices and $\gamma^5$ the 4-dimensional chirality matrix.
Imposing the chirality condition $\gamma^5 \Psi^{(4)} = +
\Psi^{(4)}$, the Dirac equation (\ref{DiracEq2}) can be written as
\begin{equation}
 \left\{ \mathrm{e}^{\frac{1}{2}cr} \hat{e}^{\mu}_{\bar{\mu}}
         \gamma^{\bar{\mu}} \hat{D}_{\mu}
        +\left( \partial_r - c
                         - \frac{1}{4}c_{1} -ie A_r(r) \right)
        +i R_0^{-1} \mathrm{e}^{\frac{1}{2}c_{1}r}
        (\partial_\theta-ie A_{\theta}(r)) \right \} \Psi^{(4)}=0. \label{DiracEq3}
\end{equation}

Now, form the equation of motion (\ref{DiracEq3}), we will search
for the solutions of the form
\begin{equation}
 \Psi^{(4)}(x,r,\theta) = \psi(x) \alpha(r) \sum \mathrm{e}^{il \theta},
\end{equation}
where $\psi(x)$ satisfies the massless 4-dimensional Dirac
equation $\hat{e}^{\mu}_{\bar{\mu}}\gamma^{\bar{\mu}}
\hat{D}_{\mu} \psi = 0$. For $s$-wave solution, Eq.
(\ref{DiracEq3}) is reduced to
\begin{eqnarray}
 \left(\partial_r - c - \frac{1}{4}c_{1}
       -ie A_r(r)
       + e R_0^{-1} \mathrm{e}^{\frac{1}{2}c_{1}r} A_{\theta}(r)
 \right) \alpha(r) = 0. \label{58}
\end{eqnarray}
The solution of this equation is given by
\begin{eqnarray}
 \alpha(r) \varpropto  {\exp}\left\{ cr + {1\over 4}c_{1}r
        +ie \int^r dr A_r(r)
        -e R_0^{-1} \int^r dr ~ \mathrm{e}^{\frac{1}{2}c_{1}r} A_{\theta}(r)
        \right\}. \label{ZeroMode}
\end{eqnarray}
So the fermionic zero mode reads
\begin{eqnarray}
\Psi \varpropto \left(%
\begin{array}{c}
  \psi \\
  0 \\
\end{array}%
\right)
 {\exp}\left\{ cr + {1\over 4}c_{1}r +ie \int^r dr A_r(r)
        -e R_0^{-1} \int^r dr ~ \mathrm{e}^{\frac{1}{2}c_{1}r} A_{\theta}(r)
        \right\}. \label{ZeroMode1}
\end{eqnarray}

Now we wish to show that this zero mode is localized on the defect
sitting around the origin $r=0$ under certain conditions. The
condition for having localized 4-dimensional fermionic field is
that $\alpha(r)$ is normalizable. It is of importance to notice
that normalizability of the ground state wave function is
equivalent to the condition that the ``coupling" constant is
nonvanishing.

Substituting the zero mode (\ref{ZeroMode1}) into the Dirac action
(\ref{DiracAction}), the effective Lagrangian for $\psi$ then
becomes
\begin{eqnarray}
 \mathcal{L}_{eff}
  &=& \int drd\theta \sqrt{-g}
      \bar{\Psi} i \Gamma^M D_M \Psi \nn\\
  &=&  I_{1/2} ~ \sqrt{-\hat{g}}~\bar{\psi}
      i \hat{e}^{\mu}_{\bar{\mu}}\gamma^{\bar{\mu}}
       \hat{D}_{\mu} \psi, \label{Leff1}
\end{eqnarray}
where
\begin{eqnarray}
 I_{1/2} &\varpropto&   \int_0^{\infty} dr
     \exp\left(\frac{1}{2} c r
       -2e R_0^{-1} \int^r dr ~ \mathrm{e}^{\frac{1}{2}c_{1}r} A_{\theta}(r)
     \right), \label{I12}
\end{eqnarray}
In order to localize spin 1/2 fermion in this framework, the
integral (\ref{I12}) should be finite. When the gauge background
vanishes, this integral is obviously divergent for $c>0$ while it
is finite for $c<0$. This situation is the same as in the case of
the domain wall in the RS framework \cite{BajcPLB2000} where for
localization of spin 1/2 field additional localization method by
Jackiw and Rebbi \cite{Jackiw} was introduced. Now let us look for
the condition for localization of spin 1/2 field. Obviously, the
$A_r$ gauge field doesn't contribute to the integral (\ref{I12}).
The requirement that the integral (\ref{I12}) should be finite is
easily satisfied. For example, a simple choice is
\begin{equation}
A_{\theta}(r) = \lambda \mathrm{e}^{-\frac{1}{2}c_{1}r},
\label{condition1}
\end{equation}
where $\lambda$ is a constant satisfying the condition
\begin{equation}
\lambda > \frac{c}{4e}R_0. \label{conditionOflambda}
\end{equation}
Another choice can be taken as the following form
\begin{equation}
A_{\theta}(r) =  \mathrm{e}^{-\frac{1}{2}c_{1}r} r^n
\label{condition2}
\end{equation}
with $n\geq 1$, or the more special and interesting form
\begin{equation}
 A_{\theta}(r) =\left(\frac{c}{4e}R_0 + r^{n} \right)
                \mathrm{e}^{-\frac{1}{2}c_{1}r}
   \label{condition3}
\end{equation}
with $n\geq 0$. So spin 1/2 field is localized on a defect with
the exponentially decreasing warp factor under condition
(\ref{condition1}) or (\ref{condition2}) or (\ref{condition3}). Of
course, there are many other choices which result in finite
$I_{1/2}$.

\subsection{Spin 3/2 fermionic field}

Next we turn to spin 3/2 field, in other words, the gravitino. Let
us start by considering the action of the Rarita-Schwinger
gravitino field:
\begin{eqnarray}
S_m = \int d^D x \sqrt{-g} \bar{\Psi}_M i \Gamma^{[M} \Gamma^N
\Gamma^{R]} D_N \Psi_R, \label{GravitinoAction}
\end{eqnarray}
where the square bracket denotes the anti-symmetrization, and the
covariant derivative is defined with the affine connection
$\Gamma^R_{MN} = e^R_{\bar{M}}(\partial_M e_N^{\bar{M}} +
\omega_M^{\bar{M} \bar{N}} e_{N {\bar{N}}})$ by
\begin{eqnarray}
 D_M \Psi_N = \partial_M \Psi_N - \Gamma^R_{MN} \Psi_R
            + \omega_M \Psi_N -ie A_M \Psi_N. \label{64}
\end{eqnarray}
From the action (\ref{GravitinoAction}), the equations of motion
for the Rarita-Schwinger gravitino field are given by
\begin{eqnarray}
\Gamma^{[M} \Gamma^N \Gamma^{R]} D_N \Psi_R = 0.
\label{GravitinoEq}
\end{eqnarray}

For simplicity, from now on we limit ourselves to the flat brane
geometry $\hat{g}_{\mu\nu} = \eta_{\mu\nu}$. After taking the
gauge condition $\Psi_r =\Psi_\theta = 0$, the non-vanishing
components of the covariant derivative are calculated as follows:
\begin{eqnarray}
 D_\mu \Psi_\nu &=& \partial_\mu \Psi_\nu
     + \frac{1}{4} c \Gamma_r \Gamma_\mu \Psi_\nu
     - ie A_{\mu} \Psi_{\nu}, \label{derivative1} \\
 D_\mu \Psi_r &=& \frac{1}{2} c\Psi_\mu, \label{derivative2}\\
 D_r \Psi_\mu &=& \partial_r \Psi_\mu
     + \frac{1}{2} c\Psi_\mu
     - ie A_{r} \Psi_{\mu}, \label{derivative3}  \\
 D_\theta \Psi_\mu &=& \partial_\theta \Psi_\mu
     + \frac{1}{4} c_1 \Gamma_r \Gamma_\theta \Psi_\mu
     - ie A_{\theta} \Psi_{\mu}. \label{derivative4}
\end{eqnarray}
Again we assume $A_{\mu}=A_{\mu}(x)$ and
$A_{r,\theta}=A_{r,\theta}(r)$, and represent $\Psi_\mu$ as the
following form
\begin{equation}
 \Psi_{\mu}=\left(%
\begin{array}{c}
  \Psi_{\mu}^{(4)} \\
  0 \\
\end{array}%
\right),\label{Psimu}
\end{equation}
where $\Psi_{\mu}^{(4)}$ is the 4D Rarita-Schwinger gravitino
field.

Imposing the chirality condition $\gamma^5 \Psi_{\mu}^{(4)} = +
\Psi_{\mu}^{(4)}$, and substituting Eqs.
(\ref{derivative1})-(\ref{Psimu}) into the equations of motion
(\ref{GravitinoEq}), we will look for the solutions of the form
\begin{eqnarray}
 \Psi_\mu^{(4)}(x,r,\theta) = \psi_\mu(x) u(r) \sum \mathrm{e}^{il \theta},
\end{eqnarray}
where $\psi_\mu(x)$ satisfies the following 4-dimensional
equations $\gamma^\mu \psi_\mu = \partial^\mu \psi_\mu =
\gamma^{[\mu} \gamma^\nu \gamma^{\rho]} (\partial_\nu -ie
A_\nu)\psi_\rho = 0$. Then the equations of motion
(\ref{GravitinoEq}) reduce to
\begin{eqnarray}
 \left(\partial_r -  \frac{1}{2} c - \frac{1}{4} c_1 -ie A_r(r)
       + e R_0^{-1} \mathrm{e}^{\frac{1}{2}c_{1}r} A_{\theta}(r)
 \right) u(r) = 0,
\end{eqnarray}
form which $u(r)$ is easily solved to be
\begin{eqnarray}
 u(r) \varpropto ~ {\exp}
      \left\{ \frac{1}{2} cr + {1\over 4}c_{1}r
        + ie \int^r dr A_r(r)
        - e R_0^{-1} \int^r dr ~ \mathrm{e}^{\frac{1}{2}c_{1}r} A_{\theta}(r)
      \right\}. \label{ZeroMode2}
\end{eqnarray}
In the above we have considered the $s$-wave solution.

Let us substitute the zero mode (\ref{ZeroMode2}) into the
Rarita-Schwinger action (\ref{GravitinoAction}). It turns out that
the effective Lagrangian becomes
\begin{eqnarray}
 \mathcal{L}_{eff}
   &=& \int drd\theta \sqrt{-g} \bar{\Psi}_M i \Gamma^{[M}
       \Gamma^N \Gamma^{R]} D_N \Psi_R \nn\\
   &=& I_{3/2} ~\bar{\psi}_\mu i \gamma^{[\mu} \gamma^\nu
       \gamma^{\rho]} (\partial_\nu -ie A_\nu)\psi_\rho. \label{L32}
\end{eqnarray}
where the integral $I_{3/2}$ is defined as
\begin{eqnarray}
 I_{3/2} &\varpropto&   \int_0^{\infty} dr \exp
    \left(\frac{1}{2} c r  -2 e R_0^{-1} \int^r dr ~
       \mathrm{e}^{\frac{1}{2}c_{1}r} A_{\theta}(r)
     \right). \label{I32}
\end{eqnarray}
In order to localize spin 3/2 fermion, the integral $I_{3/2}$ must
be finite. But this expression is equivalent to $I_{1/2}$ up to an
overall constant factor so we encounter the same result as in spin
1/2 field. This shows that the solution (\ref{ZeroMode2}) is
normalizable under the condition (\ref{condition1})  or
(\ref{condition2}) or (\ref{condition3}) for not only the
exponentially increasing but also the exponentially decreasing
warp factor.

\section{Discussions}

In this paper, we have investigated the possibility of localizing
the spin 1/2 and 3/2 fermionic fields on a brane with the
exponentially decreasing warp factor, which also localizes the
graviton. We first give a brief review of a string-like defect
solution to Einstein's equations with sources, then check
localization of fermionic fields on such a string-like defect with
the background of gauge field from the viewpoint of field theory.
We find that there is a same solution for subspace $K=D_2$-torus
with any $D_2$ and $K=D_2$-sphere with $D_2\geq 2$. It has been
found that spin 1/2 and 3/2 fields can be localized on a defect
with the exponentially decreasing warp factor if gauge and
gravitational backgrounds are considered.

Localizing the fermionic degrees of freedom on the brane or the
defect requires us to introduce other interactions but gravity.
Recently, Parameswaran {\em et al} study fluctuations about
axisymmetric warped brane solutions in 6-Dimensional minimal
gauged supergravity and proved that, not only gravity, but
Standard Model fields could ¡®feel¡¯ the extent of large extra
dimensions, and still be described by an effective 4-Dimensional
theory \cite{Parameswaran0608074}. Moreover, there are some other
backgrounds could be considered besides gauge field and
supergravity \cite{Mario}, for example, vortex background
\cite{LiuVortexFermion}. The topological vortex (especially
Abrikosov-Nielsen-Olesen vortex) coupled to fermions may lead to
chiral fermionic zero modes \cite{JackiwRossiNPB1981}. Usually the
number of the zero modes coincides with the topological number,
that is, with the magnetic flux of the vortex. In future, we wish
to extend the present work to the Abelian Higgs model.

\section*{Acknowledgement}
It is a pleasure to thank Dr Shaofeng Wu for interesting
discussions. This work was supported by the National Natural
Science Foundation of the People's Republic of China and the
Fundamental Research Fund for Physics and Mathematic of Lanzhou
University.

\end{document}